\documentclass[a4paper]{jpconf}
\usepackage{graphicx}

\newcommand{\jpsi}{${\rm J}/\psi$}
\newcommand{\lc}{$\Lambda_c^+$}
\newcommand{\btopsi}{${\rm B} \rightarrow {\rm J}/\psi$}
\newcommand{\RAA}  {\ensuremath{R_{\rm AA}}}
\newcommand{\RpA}  {\ensuremath{R_{\rm pA}}}
\newcommand{\gevc}       {\ensuremath{\mathrm{GeV}\!/c}}
\newcommand{\gevcc}       {\ensuremath{\mathrm{GeV}\!/c^2}}
\newcommand{\lqcd}{\ensuremath{\Lambda_{\rm QCD}}}

\begin{document}
\title{Experimental Overview of Open Heavy Flavor}

\author{Kai Schweda}

\address{Research Division, GSI Helmholtzzentrum f\"ur Schwerionenforschung, Darmstadt, Germany}

\ead{kschweda@cern.ch}

\begin{abstract}
These are the proceedings of the experimental overview of the production of open heavy flavor at the international conference {\it Strangeness in Quark Matter 2016}. Instead of a comprehensive overview, I focus on a few topics which the reader might find particularly interesting. 
\end{abstract}

\section{Introduction}
Heavy-flavor particles containing a heavy quark (charm or beauty) are unique probes for studies into QGP bulk properties. Their heavy masses constitute a new scale that
is much larger than the QCD scale, $m_c, m_b \gg \lqcd$, making their production cross sections at least in principle accessible to calculations in perturbative QCD. 
Their masses are also much larger than the maximum initial QGP temperature.  
Heavy quarks acquire their mass almost entirely from
the electroweak sector due to their
coupling to the Higgs field. Therefore, they remain heavy even when chiral symmetry is at least partially restored in a QGP. This makes heavy quarks a calibrated probe.
Heavy-flavor particles provide access to the degree of thermalization among quarks and gluons in the QGP. 
Due to their relatively long lifetime, heavy-flavor particles can be identified through their decay topology with the decay vertex being displaced from the collision vertex. 
\section{Results}
The Relativistic Heavy Ion Collider (RHIC) had just completed run 16 on the very day when this presentation was given. Meanwhile, an impressive data set has been collected ranging from polarized protons at up to 510 GeV and systems as large as uranium-uranium collisions. 
As part of the beam energy scan program, the center-of-mass energy was taken as low as 7 GeV. 
The PHENIX Collaboration has upgraded its detector with vertexing capabilities at mid- and forward rapidity, enabling the separation of electrons stemming from the semi-leptonic decays of beauty and charm quarks~\cite{PHENIX}. A ten-fold increased data set is presently being analyzed, which will extend the momentum reach substantially and decrease uncertainties in order to look for potential differences in suppression of beauty and charm quarks in the electron decay channel. The PHENIX Collaboration has also separated prompt \jpsi\ production from \jpsi\ production stemming from the decay of beauty hadrons, with the latter being displaced from the primary vertex. The somewhat increased fraction of \btopsi\ decays in the overall \jpsi\ production in Au--Cu collisions when compared to Au--Au collisions is attributed to a smaller level of suppression of beauty hadrons in Au--Au collisions than for \jpsi\ mesons. 

\begin{figure}[h]
\begin{minipage}{18pc}
\includegraphics[width=14pc]{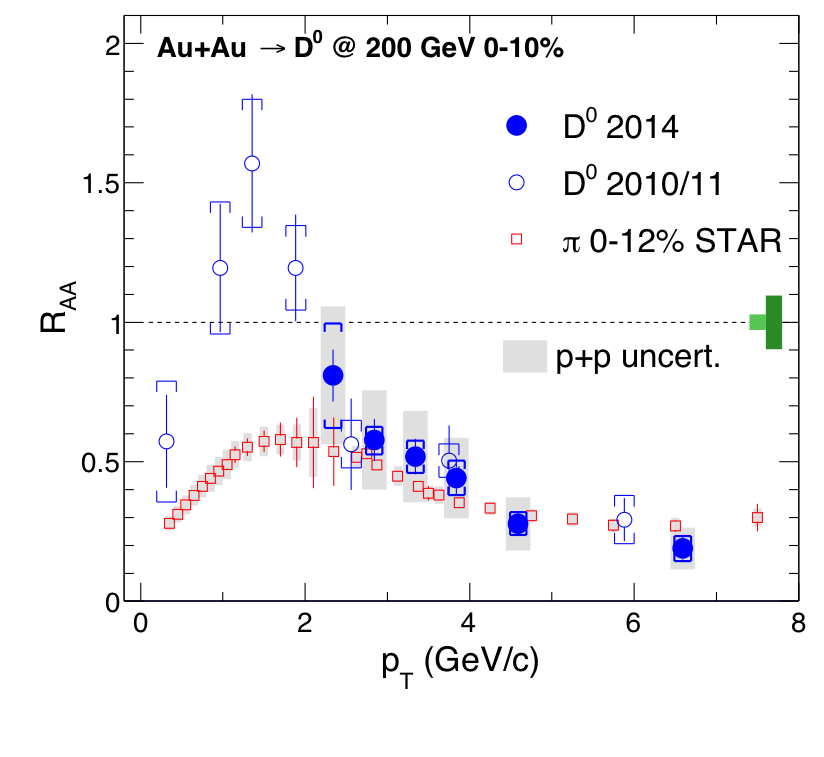}
\caption{\label{star-d0}Nuclear modification factor \RAA\ of D$^0$ mesons in Au--Au collisions at RHIC.} 
\end{minipage}\hspace{2pc}%
\begin{minipage}{18pc}
\includegraphics[width=14pc]{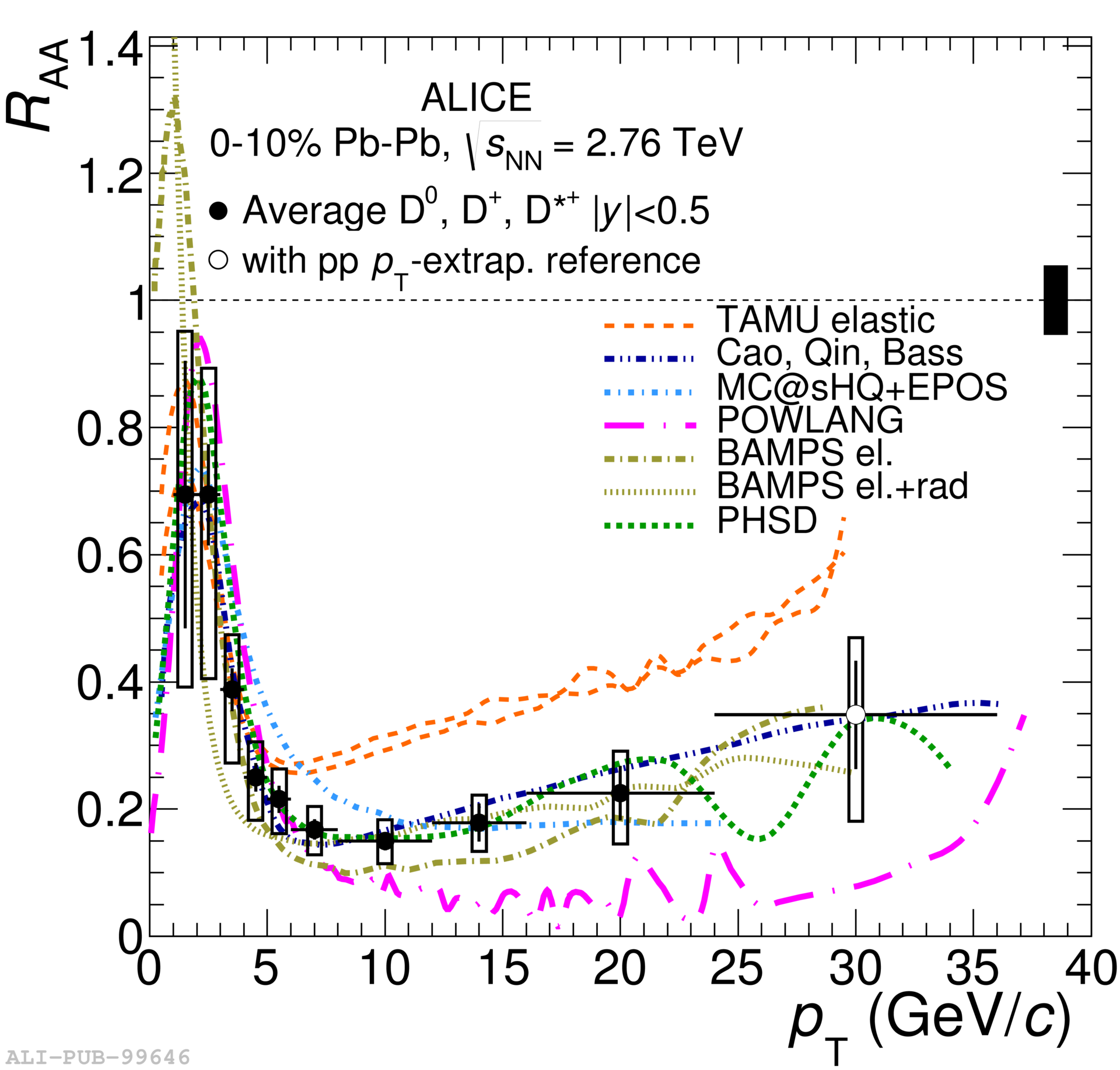}
\vspace{-2mm}
\caption{\label{alice-d}Nuclear modification factor \RAA\ of D mesons in Pb--Pb collisions at the LHC.}
\end{minipage} 
\end{figure}

The STAR Collaboration is taking advantage of its enhanced vertexing capabilities by means of the Heavy-Flavor Tracker, which is based on monolithic active pixel sensor technology~\cite{STAR}. STAR has recorded some 3 billion Au--Au collisions at 200 GeV, and as of this conference a subset (1/4) of the data has been analyzed. The nuclear modification factor in Au--Au collisions at 200 GeV shows increasingly large suppression of neutral D mesons above 2 \gevc, while around 1.5 \gevc, a characteristic bump arises with values above unity, see Fig.~\ref{star-d0}. Since charm is conserved throughout the evolution of the system, the depletion of the spectrum at large momentum should correspond to an enhancement at lower momentum. In addition, if charm participates in the collective expansion of the medium, charm flow would lead to a depopulation at very low momentum and further enhance the spectrum at intermediate momentum. Both effects would lead to a nuclear modification factor above unity, with a maximum of \RAA\ = 1.2--1.5 around 1.5 \gevc, as experimentally observed. In order to simultaneously describe the nuclear modification factor and elliptic flow measurement of D mesons at top RHIC energies, model calculations have to take into account charm diffusion with values of $(2\pi T) D$ between 2--11. These values coincide with results from the lattice.

All four large-scale LHC experiments participate in the open heavy-flavor program. This offers the unique opportunity to measure heavy quark production over the entire transverse momentum range from 0 up to 400 \gevc\ over a large rapidity range of more than seven units.  At low momentum, where most of the production cross sections is located, particle identification of decay daughters is key since topological selection is limited. This is especially true for detecting the short-lived charmed baryon \lc. 

At LHC energies, the ALICE Collaboration has pushed the measurement of D meson production to zero momentum~\cite{ALICE} in pp and p--Pb collisions. At mid-rapidity, these D mesons are virtually at rest and topological selection by means of a displaced vertex measurement is no longer applicable. 
Rather than using vertexing at these ultra-low momenta, ALICE  extracts the D meson yield by simply subtracting the large combinatorial background in the invariant mass spectrum of identified pions and kaons. This provides a precise experimental reference of the charm production cross section at mid-rapidity without any model dependence, which serves as an essential input to model calculations describing charmonium production at the phase boundary as a signal for deconfinement. Further measurements are also made in Pb--Pb collisions.

\begin{figure}[h]
\begin{minipage}{18pc} 
\includegraphics[width=14pc]{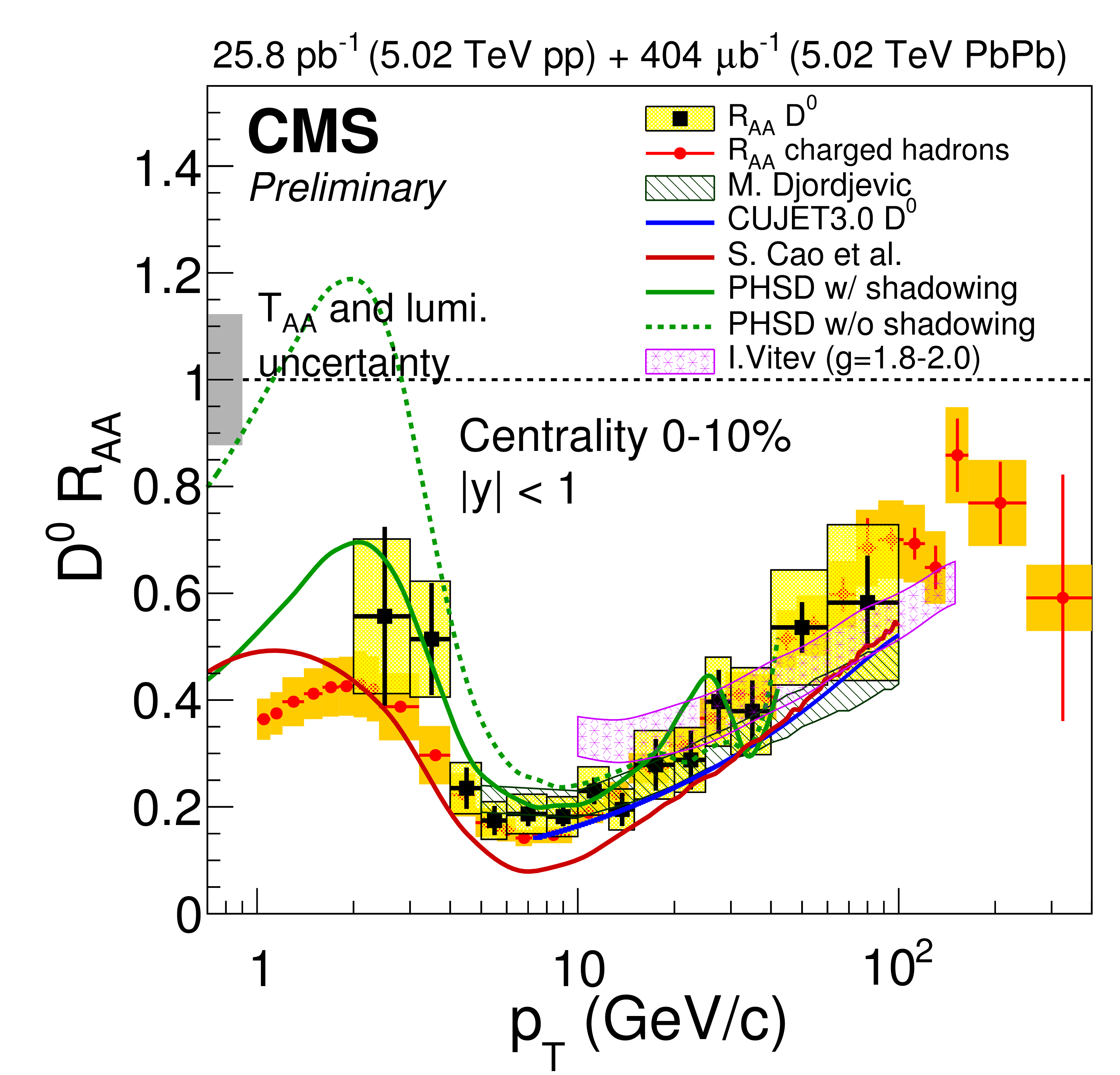}
\caption{\label{cms-d0}Nuclear modification factor \RAA\ of D$^0$ mesons in Pb--Pb collisions at the LHC.}
\end{minipage}\hspace{2pc}%
\begin{minipage}{18pc}
\includegraphics[width=14pc]{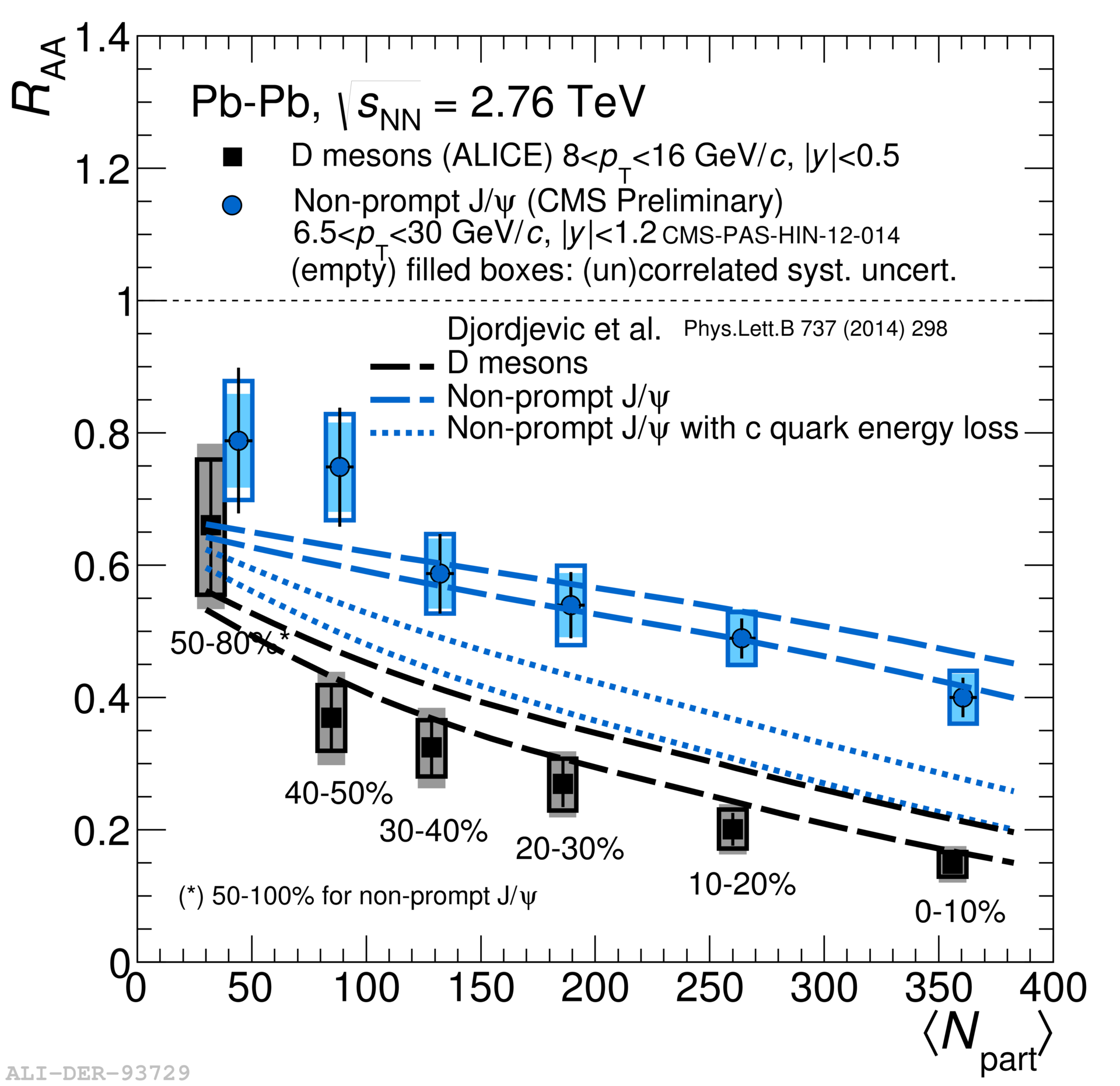}
\caption{\label{raa-db} \RAA\ of D mesons and non-prompt \jpsi\ mesons versus centrality.}
\end{minipage} 
\end{figure}

In contrast to RHIC, at LHC energies the nuclear modification factor of neutral D mesons remains substantially below unity at all measured momenta, see Fig.~\ref{alice-d}. Assuming conservation of charm in the system, i.e. charm quarks only shifting in momentum due to interactions in the medium such as energy loss processes and participating in collective flow, the initial production of charm in Pb--Pb collisions at LHC must be smaller than in pp collisions at the same energy. Transport models assume a reduction in the gluon density in each lead nucleus of up to 20\%, resulting in an initial suppression factor of 0.64 in Pb--Pb collisions. Various microscopic models are more or less able to describe the observed modification factor. While the \RAA\ values vary drastically between different models, none of them is able to consistently describe the nuclear modification factor for 
light- and heavy-flavor hadrons. 

The nuclear modification factor of neutral D mesons has been measured up to 100 \gevc\ by the CMS Collaboration~\cite{CMS}, see Fig.~\ref{cms-d0}.  At low momentum, this measurement is restricted to above 2 \gevc. In the absence of particle identification, all pairs of opposite charge have to be considered as D meson decay candidates, resulting in an overwhelming combinatorial background at low momentum. Additionally, decay daughters from a D meson decay are identified as an anti-D meson when a kaon is mistaken for a pion and vice versa,  leading to an irreducible correlated background. Note that the pp reference was recorded over 3 days of data taking, with $10^{12}$ minimum bias pp collisions being inspected. This large statistical base may allow for other experiments with lower rate capabilities in pp collisions to use reference data from other detectors.

Despite the CMS measurements showing similar values of the \RAA\ for the 10\% most central collisions when compared to  minimum bias collisions (0-100\%), the centrality dependence of the \RAA\ cannot be excluded. If binary collision scaling holds for charm production, the 10\% most central collisions contain 40\% of all D mesons in the minimum bias sample, and thus both measurements are largely correlated. Indeed, the ALICE Collaboration has already published results showing a distinct centrality dependence.
The apparent rise in \RAA\ with transverse momentum is solely due to the shape of the reference pp spectrum~\cite{Zapp:2012ak}, and not a particular prediction of any model. 

\begin{figure}[h]
\begin{minipage}{18pc}
\includegraphics[width=14pc]{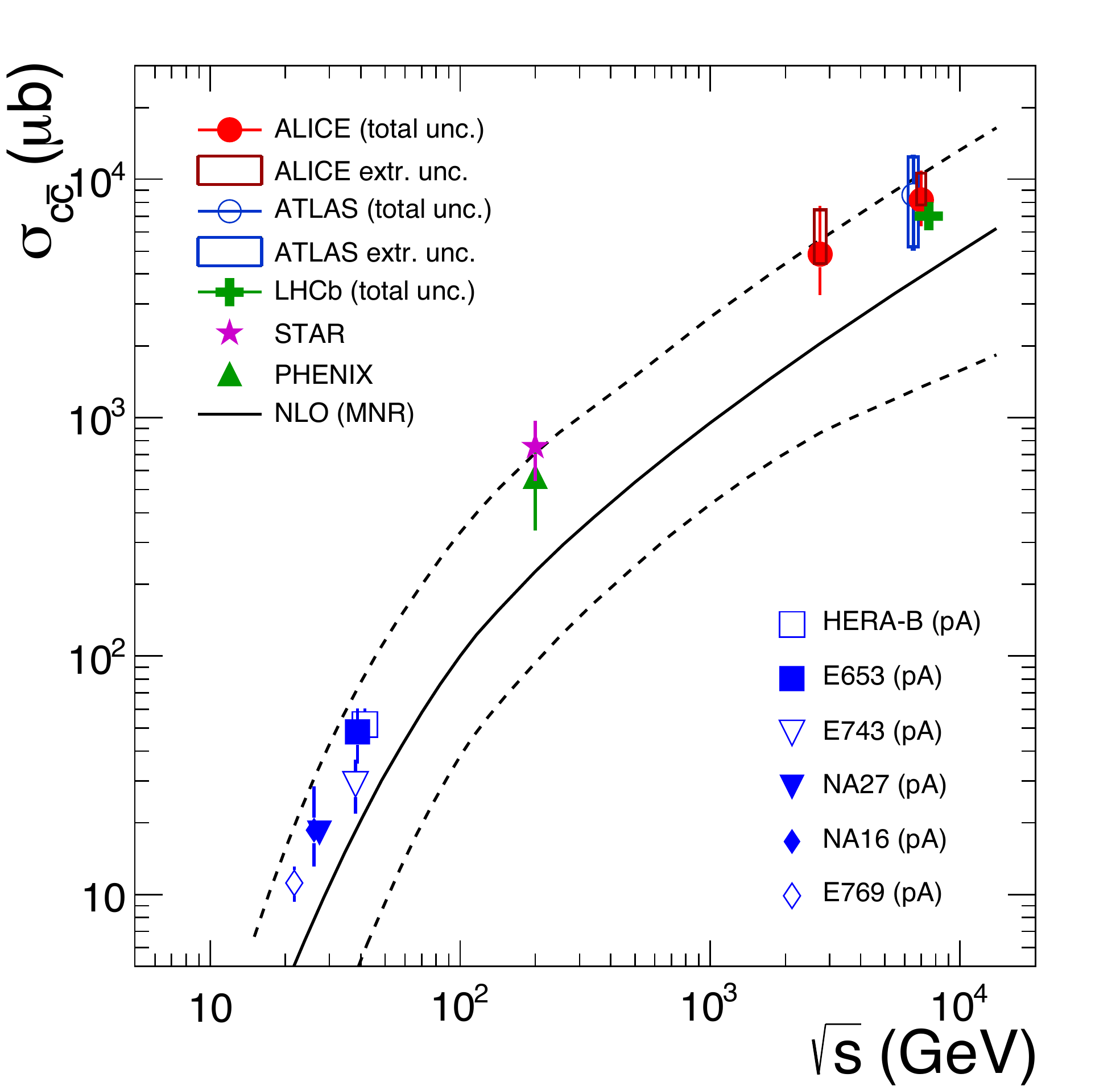}
\vspace{-1mm}
\caption{\label{sigma-ccbar} Charm production cross section versus collision energy.}
\end{minipage}\hspace{2pc}%
\begin{minipage}{18pc}
\includegraphics[trim={0 0 110cm 0}, clip, width=14pc]{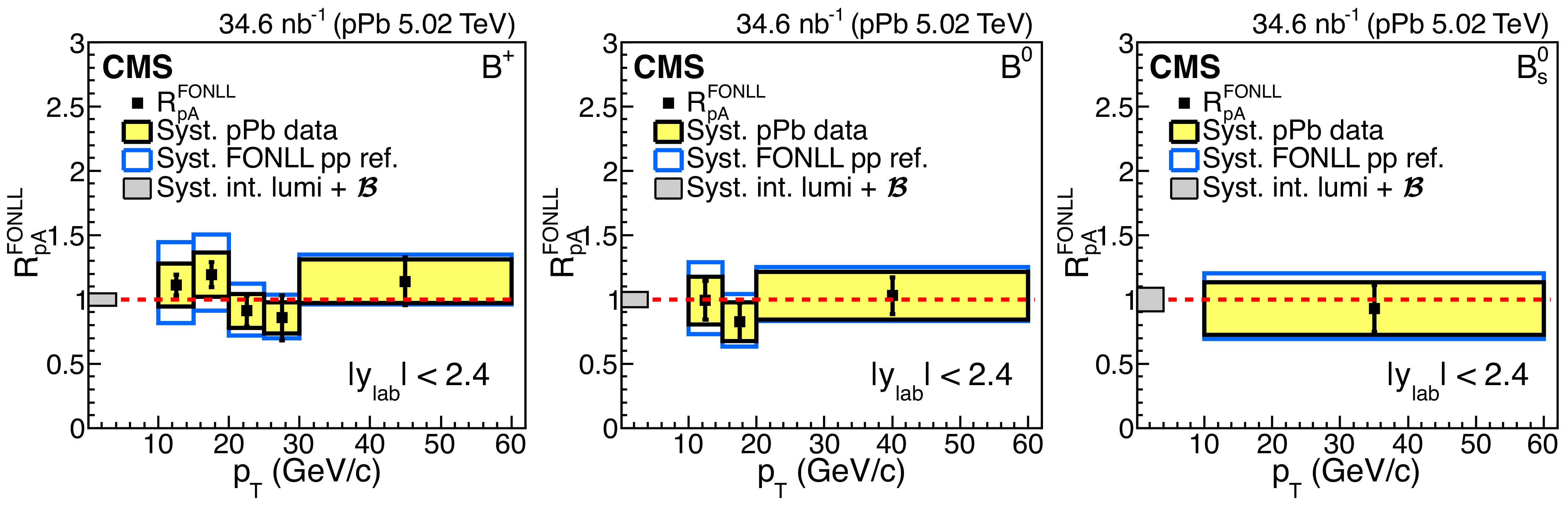}
\caption{\label{rpb-b} Nuclear modification factor \RpA\ of B$^+$ mesons in p--Pb collisions at the LHC.}
\end{minipage} 
\end{figure}

The long predicted mass dependence of beauty and charm suppression has been observed 
by comparing reconstructed D mesons from ALICE and displaced \jpsi\ mesons from B mesons in a comparable kinematic range as measured by CMS, see Fig.~\ref{raa-db}. 
As has been pointed out at this conference~\cite{gossiaux}, elastic collisions naturally lead to a mass hierarchy in suppression for simple kinematic reasons.
It would be interesting to see whether these mass differences to disappear starting at $p/m > 4$, i.e. in a momentum range above 20 \gevc, where rest masses no longer matter. In order to address the collective flow of beauty quarks, these measurements should also be extended to low momenta.

While ALICE has improved its precision in the charm production cross section by a factor two by measuring down to $p_T = 0$~\gevc, STAR has superseded its measurement on the total charm production cross section in pp collisions at 200 GeV by using run 12 data with a factor of 10 larger statistics sample, see Fig.~\ref{sigma-ccbar}. The latter value has come down by a factor of two when compared to the first publication by STAR~\cite{STAR:ccbar2004}. Interestingly, now all experimental data sit at the upper end of results in perturbative QCD at next-to-leading order within the NMR framework. 
This might hint to a smaller actual charm quark mass than the central value of $m_c$ = 1.5 \gevcc\ assumed in the calculations. A recent combined next-to-leading order QCD analysis of charm production in deep inelastic electron-proton scattering at HERA using the $\overline{\rm MS}$ running mass scheme determined a lower charm quark mass  of $m_c \approx$ 1.26 \gevcc .
A reduced charm mass would implicitly enhance the cross section of gluons splitting into pairs of charm and anti-charm quarks.  A deficit in gluon splitting processes to the production of charm has been reported at Tevatron in the associated production of Z bosons with charm quark jets and photon-tagged heavy-quark jets as well as at the LHC in the measurement of associated charm production in W final states~\cite{schweda:habil}.

Reference data from p--Pb collisions serve as a control measurement, allowing the modification of parton distribution functions in the nuclear environment to be disentangled from final-state effects in Pb--Pb collisions. The CMS Collaboration has fully reconstructed B mesons in the hadronic channel with a \jpsi\ meson in the final state, see Fig.~\ref{rpb-b}. Note, that at present these measurements are kinematically restricted above 10 \gevc\ largely due to the kinematical constraint of the muons from the \jpsi\ decay.

\begin{figure}[h]
\begin{minipage}{18pc}
\includegraphics[width=14pc]{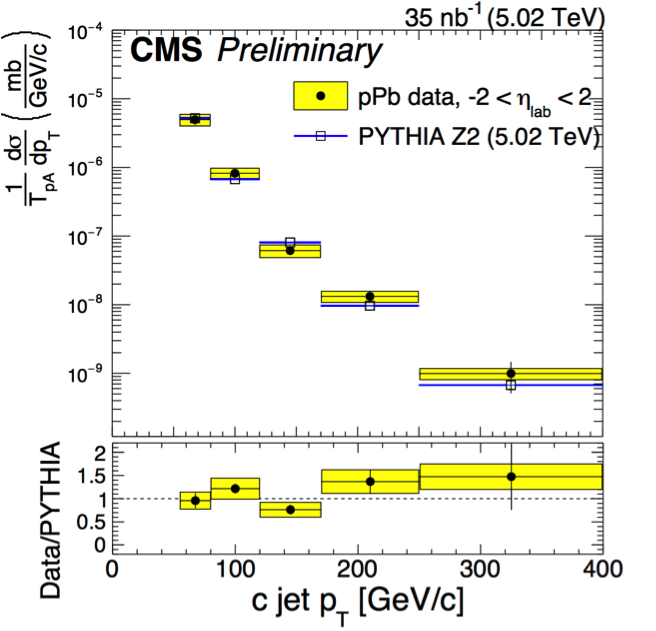}
\vspace{-2mm}
\caption{\label{pb-cjet} Charmed-jet spectrum in p--Pb collisions at LHC.}
\end{minipage}\hspace{2pc}%
\begin{minipage}{18pc}
\includegraphics[width=14pc]{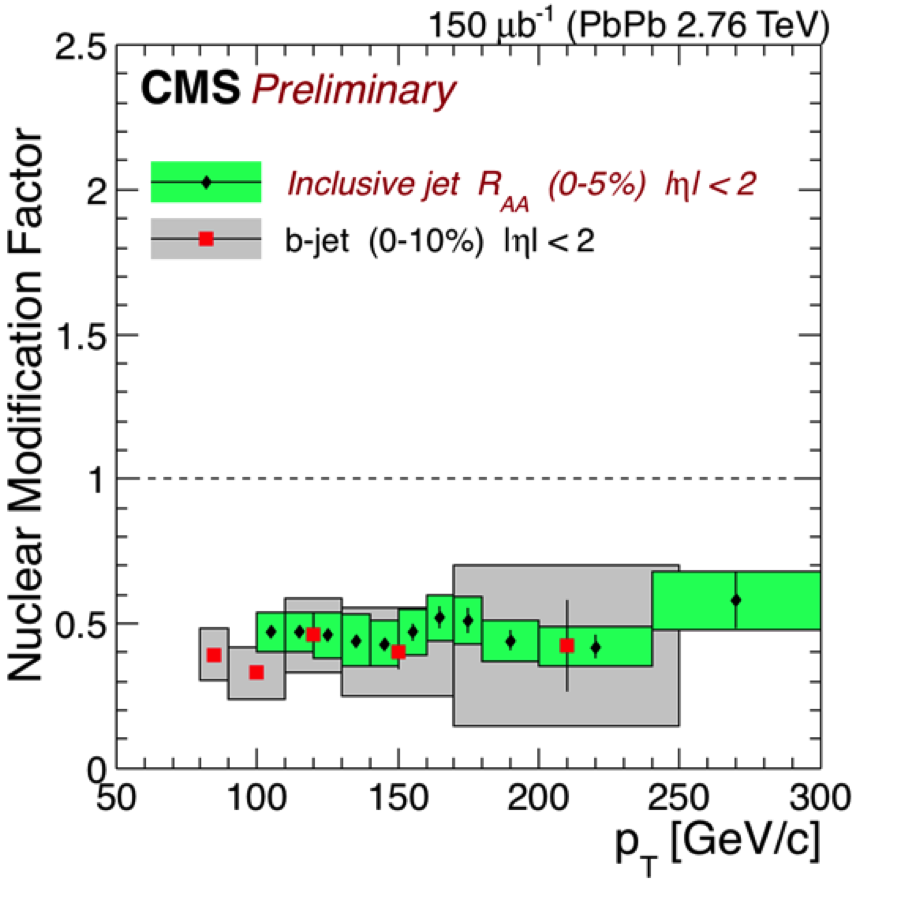}
\vspace{-3mm}
\caption{\label{raa-bjet}Nuclear modification factor \RAA\ of beauty-jets in Pb--Pb collisions at the LHC.}
\end{minipage} 
\end{figure}

Charmed jets are also now available in p--Pb collisions at the LHC, with comparisons to pQCD calculations showing no deviation within rather large systematical uncertainties, see Fig.~\ref{pb-cjet}. Overall in p--Pb collisions, the measurement of D mesons, electrons and muons from the decay of heavy quarks, and heavy-flavor jets stemming from the hadronization of charm and beauty quarks as detected by the CMS Collaboration over a large kinematical range from 50 \gevc\ up to 400 \gevc, all still leave room for nuclear modification of the initial gluon distribution of up to 20\%.
 
ALICE has measured the self-normalized yields of D mesons and electrons from semi-leptonic heavy-flavour hadron decays as a function of the charged-particle multiplicity~\cite{wagner}. The data exhibits a faster-than-linear increase with multiplicity at mid-rapidity
for both the D mesons  and the electrons from semi-leptonic heavy-flavour hadron decays. 
This might hint to the development of collectivity in high-multiplicity collisions of small systems. The seed for this collectivity seems to be related to
multi-parton interactions which become more important with increasing collision energy and are yet to be fully understood.

Beauty jets have been measured in Pb--Pb collision by the CMS Collaboration in the jet momentum range from 80 to 250 \gevc , showing
suppression by a factor of two, see Fig.~\ref{raa-bjet}. Beauty jets are identified by a secondary vertex of at least 3 charged particles which is significantly displaced from the collision vertex.
An estimate of the secondary vertex mass from reconstructed charged particles, taking missing mass into account, is used to extract the relative contributions of jet flavors, showing that a clean sample of beauty jets is identified at rather high masses. The suppression of beauty jets is comparable with that of inclusive jets, which should be dominated by gluons. With higher statistics in run 2 at the LHC, the color factor of quarks and gluons might become visible as differences in the suppression of those jets.

\begin{figure}[h]
\begin{minipage}{18pc}
\includegraphics[trim={0cm 0.0cm 10.0cm 0cm }, clip, width=14pc]{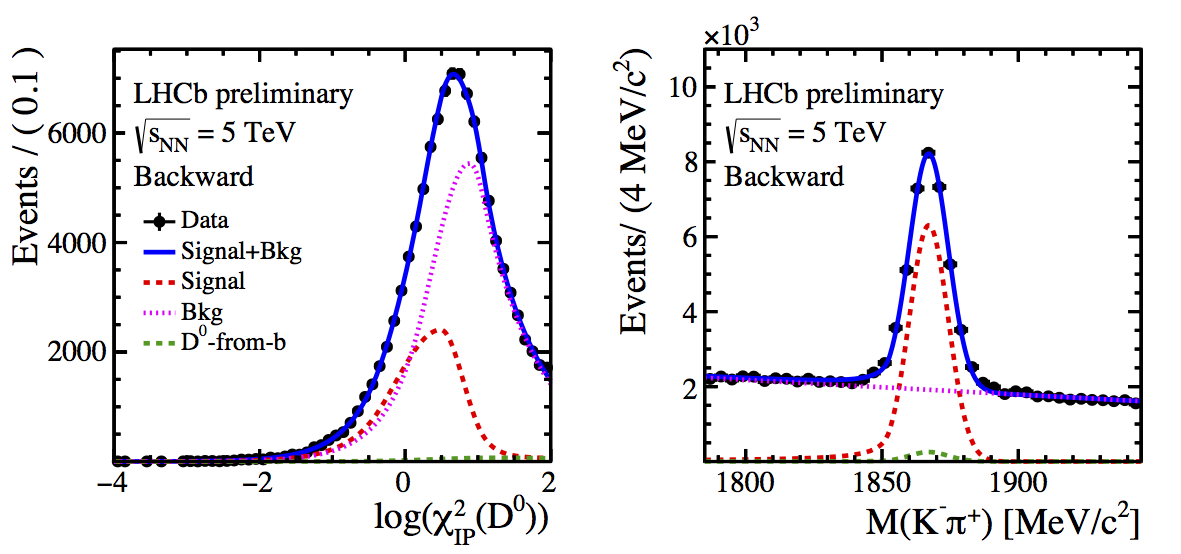}
\caption{\label{lhcb-d}Impact parameter fit of D meson candidates with LHCb.}
\end{minipage}\hspace{2pc}%
\begin{minipage}{18pc}
\includegraphics[width=14pc]{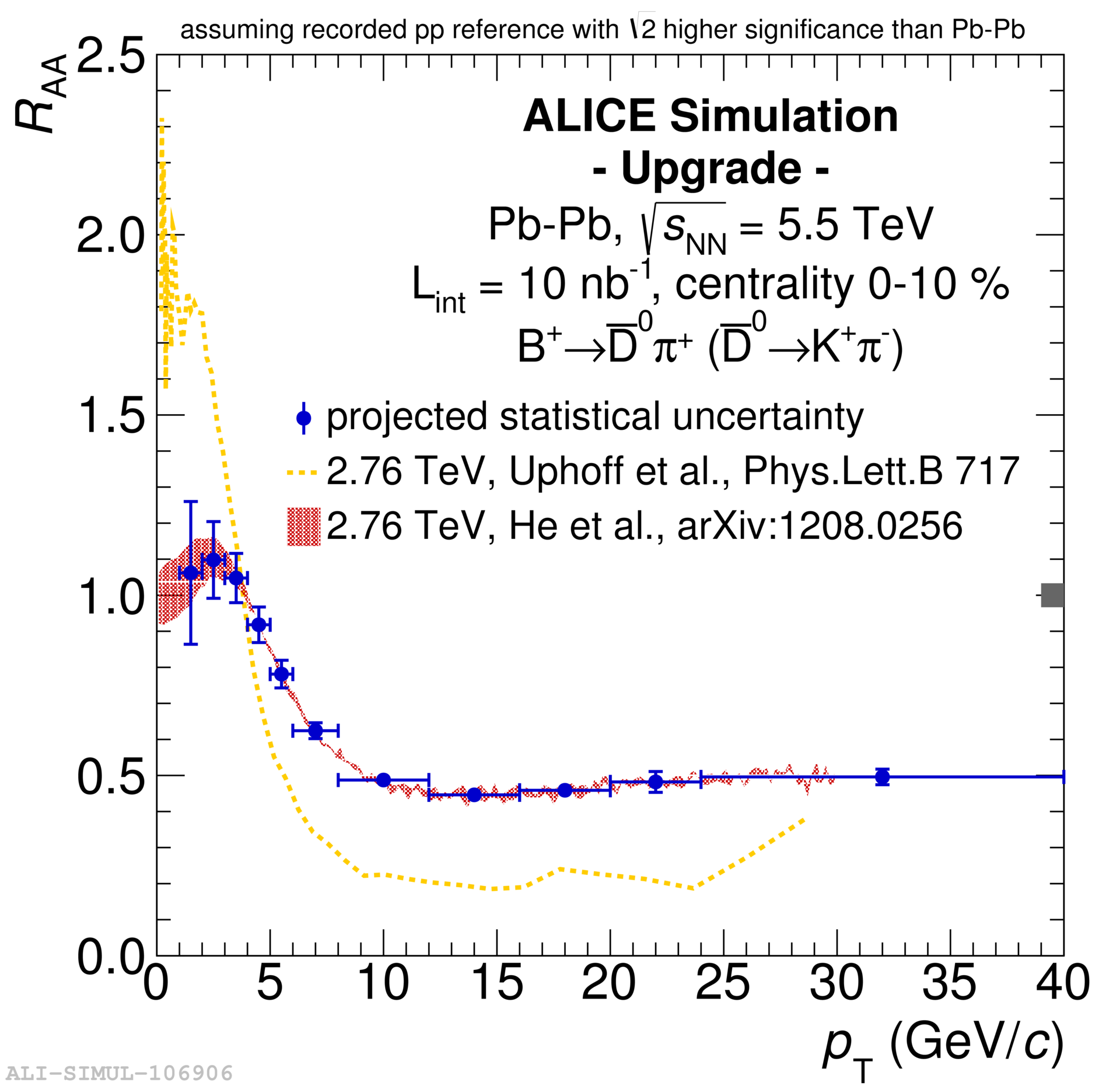}
\vspace{-2mm}
\caption{\label{raa-b}Nuclear modification factor \RAA\ of B$^+$ mesons in Pb--Pb collisions at the LHC.}
\end{minipage} 
\end{figure}

The LHCb Collaboration operated in fixed target mode by using their SMOG detector, a gas-filled device for precise luminosity measurements, as a gas target which can contain the noble gases $^4$He, $^{20}$Ne or $^{40}$Ar~\cite{LHCb}. Depending on the initial proton energy, the center of mass energy covers the SPS and RHIC energy range and converts LHCb into a mid-rapidity experiment. Due to the large boost along the beam direction, LHCb can apply precise vertexing even at vanishing transverse momentum and is thus complementary to the RHIC experiments. The feed-down contribution to D mesons stemming from the decay of beauty hadrons is experimentally disentangled by a simultaneous two-dimensional fit of the impact parameter and invariant mass, giving access to prompt open charm production without any model dependence, see Fig.~\ref{lhcb-d}.
While LHCb has participated in data taken already in run 1 of LHC, it has also recently recorded peripheral to mid-central Pb--Pb collisions.   

\section{Outlook}
Open heavy flavor measurements in ultra-relativistic heavy-ion collisions have come a long way. From the first sighting of single electrons from semi-leptonic decays and reconstructed D mesons at the Quark Matter 2004 conference at Oakland, we now have high-precision vertex capabilities, full reconstruction of D and B mesons and displaced \jpsi\ mesons,  flavor-tagged jets, correlations of heavy flavor with other particles, multiplicity dependence in pp and p--Pb collisions, etc. over an enormous range of collision energy and phase space.

The RHIC experiments are still analyzing their high-statistics data sets and all four large-scale LHC experiments are participating in the heavy-flavor program.
The near future should see detection of the elusive \lc\ baryon at both RHIC and the LHC in heavy-ion collisions, to see whether the baryon-over-meson enhancement observed in the light flavor sector is also present in the heavy flavor sector. With the factor ten increase in statistics at the ongoing run 2 at the LHC, initial state effects such as potential modifications of the gluon distribution in the environment of a lead nucleus should be constrained to significantly better than 10\%. A serious attempt must be made in order to experimentally constrain the production of charm in central lead-lead collisions to better than 10\% in order to provide essential input to all model calculations on \jpsi\ production. Since charm apparently does not fully equilibrate~\cite{greco}, it gives access to determining transport coefficients as a further characterization of the medium. This necessitates precise measurements of the centrality dependence of heavy-quark energy loss and elliptic flow, at RHIC and LHC. 
The observed mass hierarchy of heavy-quark suppression between charm and beauty should disappear above 20 \gevc, if our present understanding is correct. It remains to be seen whether the color factor appears in the different amount of suppression of gluon versus heavy-quark jets.

The ALICE Collaboration will extend full kinematic reconstruction of B mesons down to lowest momenta of 2 \gevc\ in  the decay channel
$\rm{B}^+ \rightarrow {\rm D}^0 + \pi^+$, 
see Fig.~\ref{raa-b}, after the upgrade during the second long shutdown at LHC in 2019/2020. The issue of beauty quark collectivity will be decisively addressed~\cite{silvermyer} in run 3 and 4, which is presently scheduled to start in 2021.

\subsection{Acknowledgments}
The author wishes to thank the organizers for their kind invitation and the opportunity to give a presentation at this conference and all speakers for providing their results in advance.


\end{document}